\documentclass[5p,times,twocolumn]{elsarticle}

\usepackage{setspace}
\usepackage{pgfornament}
\usepackage{wrapfig}
\usepackage{mathrsfs}  
\usepackage{subcaption}
\input{myStyle.sty}
\usepackage{eso-pic}

\newcommand{\smallMin}{\ensuremath{\scalebox{0.5}[1.0]{$-$}}}
\AddToShipoutPictureBG*{%
	\AtPageUpperLeft{%
		\setlength\unitlength{1in}%
		\hspace*{\dimexpr0.5\paperwidth\relax}
		\makebox(0,-0.75)[c]{\begin{tabular}{c c}
				Max van Haren et al,	\textit{Parameter-Varying Feedforward Control: A Kernel-Based Learning Approach,} \\
				Mechatronics 109:103337 (2025), DOI: \href{https://doi.org/10.1016/j.mechatronics.2025.103337}{10.1016/j.mechatronics.2025.103337}, uploaded May 13, 2025.  \\
		\end{tabular}}
}}

\renewcommand{\citet}{\cite}
\renewcommand{\citep}{\cite}
\newif\ifthesismode
\thesismodefalse  

\begin{document}
	\begin{frontmatter}
	\title{Parameter-Varying Feedforward Control: A Kernel-Based Learning Approach\tnoteref{label1}\tnoteref{label2}
}
	\tnotetext[label1]{This work is part of the research programme VIDI with project number 15698, which is (partly) financed by the Netherlands Organisation for Scientific Research (NWO).}
	\tnotetext[label2]{This research has received funding from the ECSEL Joint Undertaking under grant agreement 101007311 (IMOCO4.E). The Joint Undertaking receives support from the European Union Horizon 2020 research and innovation programme.}

\author[1]{Max van Haren\corref{cor1}}
\cortext[cor1]{Corresponding author.}
\ead{m.j.v.haren@tue.nl}
\author[2,1]{Lennart Blanken}
\author[1,3]{Tom Oomen}
\affiliation[1]{organization={Department of Mechanical Engineering, Control Systems Technology Section, Eindhoven University of Technology},
	addressline={Groene Loper 5},
	city={Eindhoven},
	postcode={5612 AE}, 
	country={The Netherlands}.}

\affiliation[2]{organization={Sioux Technologies},
	addressline={Esp 130},
	city={Eindhoven},
	postcode={5633 AA}, 
	country={The Netherlands}.}

\affiliation[3]{organization={Delft Center for Systems and Control, Delft University of Technology},
	addressline={Mekelweg 2},
	city={Delft},
	postcode={2628 CN}, 
	country={The Netherlands}.}

	\begin{abstract}
		The increasing demands for high accuracy in mechatronic systems necessitate the incorporation of parameter variations in feedforward control. The aim of this \manuscript is to develop a data-driven approach for direct learning of parameter-varying feedforward control to increase tracking performance. The developed approach is based on kernel-regularized function estimation in conjunction with iterative learning to directly learn parameter-varying feedforward control from data. This approach enables high tracking performance for feedforward control of linear parameter-varying dynamics, providing flexibility to varying reference tasks. The developed framework is validated on a benchmark industrial experimental setup featuring a belt-driven carriage.
\end{abstract} 
\end{frontmatter}
\section{Introduction}
Feedforward control is capable of suppressing known disturbances in motion control, specifically a reference trajectory. Feedforward control is widely applied in numerous applications, including nanopositioning \citep{Clayton2009} and robotics \citep{Grotjahn2002}. The reference tracking performance of feedforward control is directly determined by the accuracy of estimating the system's inverse dynamics \citep{Devasia2002}. As systems are designed progressively more complex, accurately describing their inverse dynamics becomes increasingly challenging.

Increasing complex dynamics in mechatronic systems can effectively be modeled using Linear Parameter-Varying (LPV) descriptions \citep{Wassink2005,Hoffmann2015}. Forward LPV models can be identified through various methods \citep{Bamieh2002,Previdi2004,Toth2010,Zhao2012}. The tracking performance of inversion-based LPV feedforward control is then directly determined by the quality of the identified forward LPV model \citep{Theis2015,DeRozario2017,Bloemers2018}. The two-step approach of forward LPV identification and inversion often degrades performance due to the intricate link between tracking performance and inverse quality, and properties such as stability are not guaranteed since inversion is typically done through optimization methods.

The limitations of inversion-based feedforward control for LPV systems have led to several approaches that directly optimize the feedforward controller based on the tracking performance. In \citet{Butcher2009}, LPV feedforward controllers are directly determined based on input-output data, while \citet{DeRozario2018a} optimizes an LPV feedforward controller using Iterative Learning Control (ILC) \citep{Bristow2006}. Both \citet{Butcher2009} and \citet{DeRozario2018a} restrict the dependence on the scheduling sequence to a generally unknown predefined structure, resulting in limited tracking performance. Furthermore, \citet{Kon2023a} employs a neural network to directly identify an LPV feedforward controller. However, its practical applicability remains limited due to the added complexity of estimating the zero dynamics of LPV systems and since the neural network is not directly capable of utilizing physical insights, which can improve estimation quality. 
Finally, direct data-driven control approaches for LPV or linear time-varying systems, such as \citep{Formentin2016,Verhoek2021,Nortmann2023}, generally do not allow for the incorporation of physical insights, and their computational complexity limits their practical applicability.
Overall, the applicability of current approaches for direct optimization of LPV feedforward controllers based on tracking performance is limited, as the structure of the dependency on the scheduling sequence must be known in advance, and no physical insights can be utilized. 

Although several techniques have been developed for LPV feedforward control, there is currently no method that directly optimizes the tracking performance without constraining the dependence on the scheduling sequence, while also allowing for the incorporation of physical insights. In this \manuscript, feedforward parameter functions are identified through kernel-regularized function estimation \citep{Pillonetto2014}, and the estimates are refined by iteratively minimizing the tracking error. Kernel methods have been successfully applied in control, including system identification \citep{Pillonetto2014,Chen2012,Pillonetto2016}, learning control \citep{Liu2018,Koller2018}, and feedforward control \citep{DeKruif2001,Blanken2020}. The main advantage of kernel-regularized function estimation is that it does not restrict the estimated function to a specific structure, but is searched over a possibly infinite-dimensional functional space that admits a finite-dimensional representation with a closed-form solution \citep{Pillonetto2014}. Furthermore, kernel-regularized estimation makes it particularly easy to incorporate physical insights of the system by enforcing high-level properties of the estimated function, such as periodicity. Additionally, the iterative nature enhances estimation quality, improving tracking performance. The following contributions are distinguished.
\begin{itemize}
	\item[C1)] Iteratively learning the feedforward parameter functions, which enhances estimation quality and improves tracking performance.
	\item[C2)] Identifying feedforward parameter functions through kernel-regularized estimation, allowing any dependence on the scheduling sequence to be modeled.
	\item[C3)] Experimental characterization and validation of the developed framework for a low-cost belt-driven carriage, which exhibits position-dependent behavior.
\end{itemize}
This work extends earlier research presented in \citet{VanHaren2022a,VanHaren2023b} by generalizing these studies to iteratively learn continuous parameter variations and apply a generic parameterization in the feedforward controller, including experimental validation.

\paragraph*{Notation} 
The discrete-time index is denoted by $\dt\in\mathbb{Z}_{[0,N-1]}$. The amount of samples in a measurement period is equal to $N$. Scalars and row and column vectors are denoted by lowercase letters, e.g., $x$. Matrices are denoted by capitals, e.g., $X$. Functions and time-dependent signals are denoted explicitly, e.g., $x(\dt)$. Time-dependent signals are vectorized as
\begin{equation}
	\label{LPVILC:eq:vectorization}
	\begin{aligned}
		x = \begin{bmatrix}
			x^\top(0) & x^\top(1) & \ldots & x^\top(N-1)	
		\end{bmatrix}^\top.
	\end{aligned}
\end{equation}
Systems are denoted calligraphically, e.g., $\mathcal{H}$. LPV systems are described using the discrete-time state-space
\begin{equation}
	\label{LPVILC:eq:LPVSS}
	\begin{aligned}
		\mathcal{H}\left(\rho\right):\quad\begin{cases}
			x(\dt+1) &= A(\rho(\dt))x(\dt) + B(\rho(\dt))u(\dt), \\
			y(\dt) &= C(\rho(\dt))x(\dt) + D(\rho(\dt))u(\dt),
		\end{cases}
	\end{aligned}
\end{equation}
with scheduling sequence $\rho(\dt)$. The response of system $\mathcal{H}(\rho)$ to input $u(\dt)$ is denoted with $y(\dt)=\mathcal{H}(\rho)u(\dt)$. Let $x(\dt)=0$ for $\dt=0$, and $u(\dt)=0$ for $\dt < 0$ and $\dt \geq N$, to obtain the finite-time LPV convolution
\begin{equation}
	\label{LPVILC:eq:LPVConvolution}
	\begin{aligned}
		\resizebox{0.91\linewidth}{!}{$ \displaystyle
		\underbrace{\begin{bmatrix}
			y(0) \\
			y(1) \\
			\vdots \\
			y(N\smallMin1)
			\end{bmatrix}}_y\!=\!\underbrace{\begin{bmatrix}
			D\left( \rho\left( 0\right) \right)  & 0 &\!\!\!\! \cdots\!\!\!\!\! & 0 \\
			C\left( \rho\left(1\right) \right) B\left( \rho\left(0\right) \right) & D\left( \rho\left(1\right) \right)  & \!\!\!\!\cdots\!\!\!\!\! & 0\\
			\vdots & \vdots & \!\!\!\!\ddots\!\!\!\!\! & \vdots \\
			C\left( \rho\left(N\smallMin1\right) \right)\!\prod_{\dt=1}^{N\smallMin2} A\left( \rho\left(\dt\right) \right)\! B\left( \rho\left(0\right) \right)\!\!\!\!\!\!\!\! & \cdots & \!\!\!\!\cdots \!\!\!\! \!& D\left( \rho\left( N\smallMin1\right) \right) 
		\end{bmatrix}}_{H} \underbrace{
		\begin{bmatrix}
			u(0) \\
			u(1) \\
			\vdots \\
		u(N\smallMin1)
		\end{bmatrix}}_u,
		$}
		\end{aligned}
		\end{equation}
		with LPV convolution matrix $H$. Linear Time-Invariant (LTI) systems are described using the forward shift operator $qu(\dt)=u(\dt+1)$.			
\section{Problem Formulation}
	In this section, a motivating application and the problem setting are shown for LPV feedforward control. Finally, the problem addressed in this \manuscript is defined.
\subsection{Motivating Application}
	\label{LPVILC:sec:MotivatingApplication}
\begin{figure}[tb]
		\centering
		\begin{subfigure}[tb]{\linewidth}
			\centering
				\includegraphics[width=\linewidth]{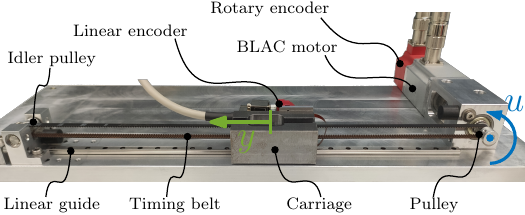}
			\caption{Photograph of experimental setup.}
			\label{LPVILC:fig:ExperimentalSystemPicture}
		\end{subfigure}
		\begin{subfigure}[tb]{\linewidth}
			\centering
			\includegraphics[width=\linewidth]{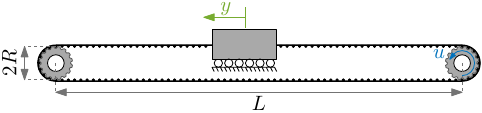}
			\caption{Sketch of experimental setup.}
			\label{LPVILC:fig:ExperimentalSystemSchematic}
		\end{subfigure}
		\caption{Experimental setup considered, where the position of the belt-driven carriage $y$ is to be controlled using the input to the motor $u$.}
		\label{LPVILC:fig:ExperimentalSystem}
	\end{figure}

	The problem addressed in this \manuscript is directly motivated by the belt-driven carriage in \figRef{LPVILC:fig:ExperimentalSystem}, which represents a transmission used regularly for mechatronic systems. Specifically, the belt-driven carriage in \figRef{LPVILC:fig:ExperimentalSystem} exhibits position-dependent dynamics, which are commonly observed in mechatronic systems and can be accurately modeled using LPV system descriptions.
	The objective of the setup is to accurately position the carriage in the $y$-direction by using the BrushLess AC (BLAC) motor. Belt-driven motion systems are typically used in applications such as printing, where the accuracy during constant velocity motion is important.	
	 The BLAC motor is connected to the carriage via a toothed pulley and a timing belt. The belt-driven carriage suffers from position-dependent dynamics due to several effects, including pulley out-of-roundness, pulley-teeth interactions, motor cogging, misalignment of the linear guide, and position-dependent drivetrain stiffness \citep[Section~4.2]{Perneder2012}, which is shown in \exampleRef{example:InverseLPV}. 
	\begin{example}
		\label{example:InverseLPV}
		By modeling the timing belt in \figRef{LPVILC:fig:ExperimentalSystem} as an elastic element under pretension with Young's modulus $E$ and cross-sectional area $A$, the perceived stiffness at the driven pulley is
		\begin{equation}
			\begin{aligned}
				k(\rho) = \frac{EA}{\frac{1}{2}L+\rho R} + \frac{EA}{1\frac{1}{2}L-\rho R},
			\end{aligned}
		\end{equation}
		for scheduling $\rho$ being the angular rotation of the driven pulley, resulting in quasi-LPV behavior. The system is modeled as a mass-spring system with continuous-time inverse dynamics \citep{VanHaren2023b}
		\begin{equation}
			\label{LPVILC:eq:InverseLPV}
			\begin{aligned}
				u(\ct) &= \left(mR+\frac{J}{R}\right)\frac{d^2y\left( \ct\right)}{d\ct^2}   + \frac{mJ}{R}\frac{d^2}{d\ct^2} \left(\frac{1}{k\left( \rho\left( \ct\right) \right)}\frac{d^2y\left(\ct\right)}{d\ct^2} \right) \\
				&= \sum_{i\in \{2,3,4\}} \frac{d^iy(\ct)}{d\ct^i}  \theta_i\left( \rho(\ct)\right),
			\end{aligned}
		\end{equation} 
		consisting of derivatives of the desired output $\frac{d^iy(\ct)}{d\ct^i} $ and LPV parameter functions $\theta_i\left( \rho(\ct)\right)$.
	\end{example}
	LPV descriptions of complex mechatronic systems, including belt-driven carriages such as the system in \figRef{LPVILC:fig:ExperimentalSystem}, directly motivate the development of parameter-varying feedforward control.
\subsection{Problem Setting}
\begin{figure}[tb]
		\centering
		\includegraphics{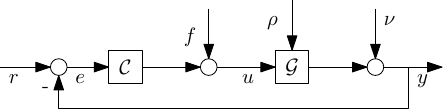}
		\caption{Control structure considered.}
		\label{LPVILC:fig:FBLayout}
	\end{figure}

	The control structure is seen in \figRef{LPVILC:fig:FBLayout}.
	The considered class of LPV systems $\mathcal{G}$ is described using the convolution in \eqref{LPVILC:eq:LPVConvolution}. The system is operating in feedback with LTI controller $\mathcal{C}$, as seen in \figRef{LPVILC:fig:FBLayout}. The tracking error, assuming zero initial state and error, is given by the interconnection of the system $\mathcal{G}$ and controller $\mathcal{C}$ by
	\begin{equation}
	\label{LPVILC:eq:error}
	e(\dt) = \mathcal{S}\left( \rho \right) \left(r(\dt) -\nu(\dt) \right)-\mathcal{J}\left(\rho\right) f(\dt), 
	\end{equation}
	with LPV sensitivity $\mathcal{S}(\rho)$ and process sensitivity $\mathcal{J}(\rho)$. LTI feedforward is not capable of compensating for the LPV dynamics in \eqref{LPVILC:eq:error}, resulting in residual error. Motivated by the inverse dynamics of LPV systems, for example \eqref{LPVILC:eq:InverseLPV} in \exampleRef{example:InverseLPV}, the goal in this \manuscript is to decrease the tracking error using the parameter-varying feedforward signal
	\begin{equation}
	\label{LPVILC:eq:FFLaw}
	f(\dt) = \psi\left(r\left( \dt\right)  \right) \theta(\rho(\dt)),
	\end{equation}
	with basis functions $\psi\left(r\left(\dt\right)\right)\in\mathbb{R}^{1\times n_\theta}$ and feedforward parameter functions $\theta(\rho(\dt))\in\mathbb{R}^{n_\theta \times 1}$. For motion systems, the basis functions $\psi\left(r\left( \dt\right) \right)$ are often chosen as derivatives of the reference signal, such as the reference velocity and acceleration \citep{Lambrechts2005,Boeren2015}. The feedforward parameterization \eqref{LPVILC:eq:FFLaw} is flexible to task variations due to the dependency on $r(\dt)$ in the basis function $\psi(r(\dt))$.
	\subsection{Problem Definition}
	The problem considered in this \manuscript is as follows. Given a reference signal $r(\dt)$, corresponding choice of basis functions $\psi\left(r\left( \dt\right)\right) \in\mathbb{R}^{1\times n_\theta}$, a model of the system $\widehat{\mathcal{G}}$, and the measured scheduling sequence $\rho(\dt)$, determine the $n_\theta$ feedforward parameter functions $\theta(\rho(\dt))\in\mathbb{R}^{n_\theta\times 1}$, such that the tracking error \eqref{LPVILC:eq:error} is minimized using the feedforward parameterization \eqref{LPVILC:eq:FFLaw}. 	
\section{Learning Feedforward Parameter Functions}
\label{LPVILC:sec:method}
In this section, feedforward parameter functions are iteratively learned through kernel-regularized estimation. 
The non-parametric nature of kernel-regularized estimation enables modeling any dependence on the scheduling sequence, and iterative learning improves estimation quality, both contributing to C1 and C2. 

Furthermore, the design of kernels and the developed procedure are presented.

\subsection{Iterative Learning of Feedforward Parameter Functions}
\label{LPVILC:sec:IterativeLearning}
The key idea is to iteratively learn the feedforward parameter functions $\theta(\rho(\dt))$ in \eqref{LPVILC:eq:FFLaw} such that the tracking error is minimized. 
The error \eqref{LPVILC:eq:error} is approximated in the next trial $j+1$ for feedforward signal \eqref{LPVILC:eq:FFLaw} based on a model of the process sensitivity $\widehat{\mathcal{J}}$ as
\begin{equation}
	\label{LPVILC:eq:predictedError}
	\begin{aligned}
		\hat{e}_{j+1}(\dt) =e_j(\dt)-\widehat{\mathcal{J}}\left(\rho_j\right)\psi\left( r_j\left( \dt\right) \right) \left( \theta_{j+1}\left( \rho_{j}\left( \dt\right) \right) -\theta_j\left( \rho_{j}\left( \dt\right) \right)\right). 
	\end{aligned}
\end{equation}
The parameters are then iteratively learned by minimizing a certain cost function with respect to the approximated error, that is
\begin{equation}
	\begin{aligned}
		\min_{\theta_{j+1}} V\left(\hat{e}_{j+1}\right).
	\end{aligned}
\end{equation}
As a result, the feedforward parameters are iteratively updated over iterations or trials $j$ as
\begin{equation}
	\label{LPVILC:eq:deltaUpdate}
	\begin{aligned}
		\theta_{j+1}(\rho^*) = \theta_{j}(\rho^*)+\theta_{j}^\Delta(\rho^*),
	\end{aligned}
\end{equation}
for arbitrary scheduling value $\rho^*$. 
\begin{remark}
	By choosing $V\left(\hat{e}_{j+1}\right) = \sum_{\dt=0}^{N-1}\hat{e}_{j+1}^2(\dt)$ and optimizing for LTI feedforward parameters $\theta_j(\rho(\dt))\equiv\bar{\theta}_j \;\forall \rho(\dt)$, LTI ILC with basis functions \citep{VanDeWijdeven2010} is recovered.
\end{remark}
The iterative procedure is effective since it directly minimizes the least-squares tracking error, and uses both a model of the process sensitivity and the measured data, thus reducing the requirements on the modeling quality \citep{VanDerMeulen2008}. 
A key challenge is to learn $\theta_{j}^\Delta(\rho^*)$ in \eqref{LPVILC:eq:deltaUpdate} without constraining the estimated function to a certain structure, which is presented in the next section.

\subsection{Kernel-Regularized Learning of Feedforward Parameter Functions}
Unlike traditional estimation methods that restrict the estimated function $\theta_{j}^\Delta(\rho^*)$ in \eqref{LPVILC:eq:deltaUpdate} to a specific structure, the function $\theta_{j}^\Delta(\rho^*)$ is estimated in a possibly infinite-dimensional function space $\mathscr{H}$ and can be evaluated at any arbitrary $\rho^*\in\mathbb{R}$. Specifically, the feedforward parameter functions are estimated by iteratively minimizing the predicted least-squares tracking error \eqref{LPVILC:eq:predictedError} with a regularization term $J$ to prevent overfitting and ill-posedness, i.e.,
\begin{equation}
	\begin{aligned}
		\min_{\theta_{j}^\Delta\in\mathscr{H}} \sum_{\dt=0}^{N-1}
		\hat{e}_{j+1}^2(\dt) + \gamma J\left( \theta_{j}^\Delta\right) .
	\end{aligned}
\end{equation}
The regularizer $J$ can be chosen to penalize unwanted behavior of the estimated functions $\theta_{j}^\Delta$. For example, the energy of the estimated functions $\theta_{j}^\Delta$ can be reduced by penalizing with $J(\theta_{j}^\Delta) =\int{\left(\theta_{j}^\Delta(\rho)\right)^\top\theta_{j}^\Delta(\rho)\;d\rho}$.

Properties of the estimated function are effectively enforced through designing $\mathscr{H}$ as a Hilbert space, and choosing the regularizer as the squared norm in this space,
\begin{equation}
	\label{LPVILC:eq:RegularizedCostFunction}
	\begin{aligned}
		\min_{\theta_{j}^\Delta\in\mathscr{H}} \sum_{\dt=0}^{N-1}
		\hat{e}_{j+1}^2(\dt) + \gamma \|\theta_{j}^\Delta\|_\mathscr{H}^2,
	\end{aligned}
\end{equation}
with Reproducing Kernel Hilbert Space (RKHS) norm $\|\theta_{j}^\Delta\|_\mathscr{H}^2=\left\langle \theta_{j}^\Delta,\theta_{j}^\Delta\right\rangle_\mathscr{H}$ \citep{Pillonetto2014}. The RKHS $\mathscr{H}$ is associated with a kernel function that is capable of reproducing every function in the space, in this case the $n_\theta\times n_\theta$ kernel function matrix
\begin{equation}
	\label{LPVILC:eq:kernelfunctionmatrix}
	K\left(\rho^*,\rho\right) = \begin{bmatrix}
			k_{11}\left(\rho^*,\rho\right) & k_{12}\left(\rho^*,\rho\right) & \cdots & k_{1n_\theta}\left(\rho^*,\rho\right) \\
			k_{21}\left(\rho^*,\rho\right) & k_{22}\left(\rho^*,\rho\right) & \cdots & k_{2 n_\theta}\left(\rho^*,\rho\right) \\
			\vdots & \vdots & \ddots & \vdots \\
			k_{n_\theta 1}\left(\rho^*,\rho\right) & k_{n_\theta 2}\left(\rho^*,\rho\right) & \cdots & k_{n_\theta n_\theta}\left(\rho^*,\rho\right) 
		\end{bmatrix},
\end{equation}
where each kernel function $k_{ij}$ describes the correlation between feedforward parameters $i$ and $j$. As a result, the kernel functions enable to enforce desired properties of the estimated function $\theta_{j}^\Delta\left(\rho^*\right)$, such as smoothness or periodicity.
\begin{remark}
	Cross-correlation between different feedforward parameters, which is commonly observed for motion systems \citep{VanHaren2023b}, is directly enabled by setting $k_{ij} \neq 0 \; \forall i\neq j$.
\end{remark}
Since the kernel-regularized estimates of the feedforward parameter functions are estimated in the possibly infinite dimensional space $\mathscr{H}$, the function estimates are not restricted to a specific class of functions and are thus capable of modeling any feedforward parameter function.

Although the feedforward parameter functions are modeled in a possibly infinite dimensional space $\mathscr{H}$, it admits a finite-dimensional solution through the representer theorem \citep[(63)]{Pillonetto2014}
\begin{equation}
	\begin{aligned}
		\label{LPVILC:eq:representerTheorem}
		\theta_j^\Delta\left( \rho^*\right)  = \sum_{\dt=0}^{N-1}  K\left( \rho^*,\rho_j(\dt)\right)  \psi^\top\left( r_j\left( \dt\right) \right)  \widehat{\mathcal{J}}\left( \rho_j\right)   \hat{c}_{j}(\dt),
	\end{aligned}
\end{equation}
where kernel function matrix $K(\rho^*,\rho_j(\dt)) \in \mathbb{R}^{n_\theta\times n_\theta}$ is determined with \eqref{LPVILC:eq:kernelfunctionmatrix}. The (modified) representers $\hat{c}_j = \begin{bmatrix}\hat{c}_{j}(0) & \hat{c}_{j}(1) &\cdots & \hat{c}_{j}(N-1)\end{bmatrix}^\top \in\mathbb{R}^N$ are given by \citep[(64) and (70b)]{Pillonetto2014}
\begin{equation}
	\label{LPVILC:eq:representerSolution}
	\hat{c}_j =  \left(\widehat{J}_j\Psi_j K_j\Psi_j^\top {\widehat{J}_j}^\top+\gamma I_N\right)^{-1}e_j,
\end{equation}
with convolution matrix $\widehat{J}_j\in\mathbb{R}^{N\times N}$ of LPV system $\widehat{\mathcal{J}}$ evaluated at the measured scheduling sequence $\rho_j$ using \eqref{LPVILC:eq:LPVConvolution}, and the kernel matrix $K_j\in \mathbb{R}^{Nn_\theta\times Nn_\theta}$ is evaluated by
\begin{equation}
	\label{LPVILC:eq:liftedKernelMatrix}
	\ifthesismode
		K_j \!= \!\begin{bmatrix}
			K\left( \rho_j(0),\rho_j(0)\right)  & K\left( \rho_j(0),\rho_j(1)\right)  & \!\!\!\cdots\!\!\! &\!\!K\left( \rho_j(0),\rho_j(N\smallMin1)\right)  \\
			K\left( \rho_j(1),\rho_j(0)\right) & K\left( \rho_j(1),\rho_j(1)\right)  & \!\!\!\cdots &\!\! K\left( \rho_j(1),\rho_j(N\smallMin1)\right)  \\
			\vdots & \vdots & \!\!\!\ddots\!\!\! & \!\!\vdots \\
			K\left( \rho_j\left( N\smallMin1\right) ,\rho_j\left( 0\right) \right) \!\!\! & K\left( \rho_j\left( N\smallMin1\right) ,\rho_j\left( 1\right) \right)  & \!\!\!\cdots\!\!\! & \!\!K\left( \rho_j\left( N\smallMin1\right) ,\rho_j\left( N\smallMin1\right) \right)  \\
		\end{bmatrix}. 
	\else
	\resizebox{\linewidth}{!}{%
			$ \displaystyle
				K_j \!= \!\begin{bmatrix}
				K\left( \rho_j(0),\rho_j(0)\right)  & K\left( \rho_j(0),\rho_j(1)\right)  & \!\!\!\cdots\!\!\! &\!\!K\left( \rho_j(0),\rho_j(N\smallMin1)\right)  \\
				K\left( \rho_j(1),\rho_j(0)\right) & K\left( \rho_j(1),\rho_j(1)\right)  & \!\!\!\cdots &\!\! K\left( \rho_j(1),\rho_j(N\smallMin1)\right)  \\
				\vdots & \vdots & \!\!\!\ddots\!\!\! & \!\!\vdots \\
				K\left( \rho_j\left( N\smallMin1\right) ,\rho_j\left( 0\right) \right) \!\!\! & K\left( \rho_j\left( N\smallMin1\right) ,\rho_j\left( 1\right) \right)  & \!\!\!\cdots\!\!\! & \!\!K\left( \rho_j\left( N\smallMin1\right) ,\rho_j\left( N\smallMin1\right) \right)  \\
			\end{bmatrix}. $}
		\fi
\end{equation}
Note that the feedforward parameters for the next trial in \eqref{LPVILC:eq:representerTheorem} are calculated based on the scheduling sequence measured in the current trial.  The basis function matrix $\Psi_j\in \mathbb{R}^{N\times Nn_\theta}$ is constructed such that $f_j = \Psi_j \theta_j$ as
\begin{equation}
	\label{LPVILC:eq:liftedBasisFunctionMatrix}
	\Psi_j = \begin{bmatrix}
	\psi\left( r_j\left(0\right) \right) & 0 & \cdots & 0 \\
	0 & \psi\left( r_j\left(1\right) \right) & \cdots & 0 \\
	\vdots &\vdots & \ddots & \vdots \\
	0 & \cdots & 0 & \psi\left( r_j\left(N-1\right) \right)
	\end{bmatrix}.
\end{equation}

The feedforward parameters for the next trial \eqref{LPVILC:eq:deltaUpdate} are estimated by propagating the feedforward update using the representer theorem \eqref{LPVILC:eq:representerTheorem} over the trials, resulting in
\begin{equation}
	\label{LPVILC:eq:ParameterUpdate}
	\begin{aligned}
		\theta_{j+1}\left( \rho^*\right)  = \sum_{i=0}^{j} \sum_{\dt=0}^{N-1} K\left( \rho^*,\rho_i(\dt)\right)\psi^\top\left( r_i\left( \dt\right) \right)  \widehat{\mathcal{J}}\left( \rho_i\right) \hat{c}_i(\dt).
	\end{aligned}
\end{equation}
Kernel-regularized learning of LPV feedforward parameter functions in \eqref{LPVILC:eq:ParameterUpdate} estimates the functions without restricting the dependence on the scheduling sequence, while allowing for the incorporation of prior knowledge.
\begin{remark}
	For the special case where the scheduling sequence during iterative learning is constant, i.e., $\rho_j(\dt)=\rho(\dt)\; \forall j$, the parameter update \eqref{LPVILC:eq:ParameterUpdate} simplifies to
	\begin{equation}
		\label{LPVILC:eq:TrialInvariantParameterUpdate}
		\begin{aligned}
			\theta_{j+1}\left( \rho^*\right)  = \theta_{j}\left( \rho^*\right)  + \sum_{\dt=0}^{N-1}  K\left( \rho^*,\rho(\dt)\right)\psi^\top\left( r_j\left( \dt\right) \right)  \widehat{\mathcal{J}}\left( \rho\right)  \hat{c}_{j}(\dt).
		\end{aligned}
	\end{equation}
\end{remark}
\begin{remark}
	The convergence of learning feedforward parameter functions \eqref{LPVILC:eq:ParameterUpdate} is primarily influenced by 
	\begin{enumerate}
	\item the quality of model $\widehat{\mathcal{J}}$;
	\item the choice of kernel functions in \eqref{LPVILC:eq:kernelfunctionmatrix}; and
	\item the regularization coefficient $\gamma$ in \eqref{LPVILC:eq:RegularizedCostFunction} and \eqref{LPVILC:eq:representerSolution}.
	\end{enumerate}
Generally, the regularization parameter $\gamma$ can be increased to ensure convergence of the framework. For a trial-invariant basis function $\Psi_j=\Psi \;\forall j$ and scheduling sequence $\rho_j=\rho \;\forall j$, which leads to $J_j=J,\: K_j=K \;\forall j$, and no modeling uncertainty $\widehat{J}=J$, the propagation of feedforward parameters \eqref{LPVILC:eq:ParameterUpdate} and \eqref{LPVILC:eq:TrialInvariantParameterUpdate} is written in vector notation $\theta = \begin{bmatrix}\theta(\rho(0)) &\theta(\rho(1)) & \cdots & \theta(\rho(N-1))\end{bmatrix}^\top$ as
	\begin{equation}
		\label{LPVILC:eq:trialInvariantConvergence}
		\begin{aligned}
			\theta_{j+1} &= \theta_{j} + K\Psi^\top{J}^\top\left({J}\Psi K\Psi^\top J^\top+\gamma I_N\right)^{-1}e_j \\
			&= \gamma\left(K\Psi^\top{J}^\top{J}\Psi+\gamma I_{Nn_\theta}\right)^{-1} \theta_j + \left(\ldots\right) S\left(r_j-\nu_j\right),
		\end{aligned}
	\end{equation}
with $e_j$ from \eqref{LPVILC:eq:error}. 
\ifthesismode
Monotonic convergence of the feedforward parameters $\theta_{j+1}$ at the trial-invariant scheduling $\rho$ \eqref{LPVILC:eq:trialInvariantConvergence} is guaranteed if \citep{VanDeWijdeven2010} \[\bar{\sigma}\left(\gamma\left(K\Psi^\top{J}^\top{J}\Psi+\gamma I_{Nn_\theta}\right)^{-1}\right) < 1,\] with $\bar{\sigma}$ the largest singular value.
\else
Monotonic convergence of the feedforward parameters $\theta_{j+1}$ at the trial-invariant scheduling $\rho$ \eqref{LPVILC:eq:trialInvariantConvergence} is guaranteed if $\bar{\sigma}\left(\gamma\left(K\Psi^\top{J}^\top{J}\Psi+\gamma I_{Nn_\theta}\right)^{-1}\right) < 1$ \citep{VanDeWijdeven2010}, with $\bar{\sigma}$ the largest singular value.
\fi
Furthermore, ILC with basis functions and its associated convergence properties \citep{VanDeWijdeven2010} are recovered by $K\left( \rho^*,\rho\right) =I$ and $\gamma=0$.
\end{remark}
\begin{remark}
	The robustness against modeling uncertainties of ILC \citep{VanDeWijdeven2010} allows the use of LTI model $\widehat{\mathcal{J}}(q)$ instead of the LPV model $\widehat{\mathcal{J}}\left(\rho_j\right)$.		
\end{remark}
\subsection{Design of Kernel Functions}
The developed update law requires the design of kernel functions $k_{ij}\left(\rho^*, \rho\right)$ in matrix $K\left(\rho^*,\rho\right)$ in \eqref{LPVILC:eq:kernelfunctionmatrix}. Many different kernel functions are possible, and the suitable choice depends on the problem at hand. Three examples of high-level properties that can be enforced on the feedforward parameters through the use of kernel functions are the following.
\begin{enumerate}
	\item Constant feedforward parameter functions are realized by the constant kernel
	\begin{equation}
		\label{LPVILC:eq:constantKernel}
		k_{ij}^c\left(\rho^*,\rho\right) = \sigma^2,
	\end{equation}
	with hyperparameter $\sigma^2$ determining the average distance of the function to its mean. 
	\begin{remark}
			\label{rem:RegularizedLTIBFILC}
		The constant kernel recovers LTI ILC with basis functions \citep{VanDeWijdeven2010}, with parameters estimated using Tikhonov regularization for $\gamma>0$ and without regularization for $\gamma=0$.
	\end{remark}
	\item Smooth feedforward parameter functions are estimated using the squared-exponential kernel
	\begin{equation}
		\label{LPVILC:eq:SE}
		k_{ij}^{SE}\left(\rho^*,\rho\right) = 
		\sigma^2 \exp\left(-\frac{\left(\rho^*-\rho\right)^2}{2\ell^2}\right),
	\end{equation}
	where $\sigma^2$ has the same function as for the kernel \eqref{LPVILC:eq:constantKernel}, and the hyperparameter $\ell$ determines the level of smoothness of the estimated function. 
	\item Periodic feedforward parameter functions are realized through the periodic kernel
	\begin{equation}
		\label{LPVILC:eq:per}
		k_{ij}^{per}\left(\rho^*,\rho\right) = \sigma^2 \exp\left(-\frac{2 \sin^2\left(\pi\left|\rho^*-\rho\right| / p\right)}{\ell^2}\right),
	\end{equation}
	where $\sigma^2$ and $\ell$ have the same role as in the squared-exponential kernel \eqref{LPVILC:eq:SE}, and the hyperparameter $p$ forces the feedforward parameter functions to be periodic with period $p$.
\end{enumerate}
In addition, multiple kernels can be combined such that they have the properties of various kernels, and can be trivially extended for multidimensional inputs \citep{Rasmussen2004}. The choice of kernel functions is determined by the situation at hand, where several guidelines are given in \citet{Pillonetto2014} and \citet[Chapter~4]{Rasmussen2004}. Considering the timing-belt system in \figRef{LPVILC:fig:ExperimentalSystem}, periodicity along the pulley circumference can be embedded in the feedforward parameter functions through the use of a periodic kernel.
\begin{remark}
	The LPV feedforward signal should in some cases be dynamically dependent, meaning that it should be dependent on derivatives or time-shifted values of $\rho$ \citep{Kon2023a,VanHaren2023b}. The kernel functions can be straightforwardly extended by using these values as additional input to the kernel.
\end{remark}

\subsection{Procedure}
The developed procedure that iteratively learns feedforward parameter functions through kernel-regularized estimation is summarized in Procedure~\ref{LPVILC:proc:1}.
\begin{figure}[H]
	\vspace{0mm}
	\hrule 
	\vspace{1mm}
	\begin{proced}[Iterative kernel-regularized learning of LPV feedforward parameters] \hfill \vspace{0.5mm} \hrule
		\label{LPVILC:proc:1}
		\vspace{0.5mm}
		\textbf{Inputs:} Model $\widehat{\mathcal{J}}$, reference signal $r_j$, choice of basis functions in $\psi$, which (measured) signals are the scheduling sequence $\rho_j$, kernel function matrix $K(\rho_j^*,\rho_j)$ from \eqref{LPVILC:eq:kernelfunctionmatrix} and corresponding hyperparameters.
		\begin{enumerate}
			\item Initialize $\theta_{0}\left( \rho^* \right) $, e.g., $\theta_{0}\left( \rho^*\right)=0 \; \forall \rho^*$.
			\item For $j \in \mathbb{Z}_{[0,N_{trial}-1]}$
			\begin{enumerate}
				\item Compute basis functions $\psi(r_j(\dt))$ in \eqref{LPVILC:eq:FFLaw} using the reference $r_j(\dt)$, and construct basis function matrix $\Psi_j$ using \eqref{LPVILC:eq:liftedBasisFunctionMatrix}.
				\item Compute $f_j(\dt)$ with \eqref{LPVILC:eq:FFLaw}.
				\item Apply $f_j(\dt)$ and $r_j(\dt)$ to the system in \figRef{LPVILC:fig:FBLayout}, and measure error $e_j(\dt)$ and scheduling sequence $\rho_j(\dt)$. 
				\item Construct convolution matrix $\widehat{J}_j$ using \eqref{LPVILC:eq:LPVConvolution} with model $\widehat{\mathcal{J}}$ and measured scheduling sequence $\rho_j(\dt)$. \label{step:LPVJ}
				\begin{itemize}
					\item If no LPV model is available, set $\widehat{\mathcal{J}}\left( \rho_j\right) $ to an LTI approximate, i.e., $\widehat{\mathcal{J}}(q)$.
				\end{itemize}
				\item Determine kernel matrix $K_j$ based on measured scheduling sequence $\rho_j$ using \eqref{LPVILC:eq:liftedKernelMatrix}.
				\item Calculate the representers $\hat{c}_j$ for trial $j$ using \eqref{LPVILC:eq:representerSolution}.
				\item Compute the feedforward parameters for the next trial $j+1$ based on the current measured scheduling sequence $\rho_j$, meaning $\theta_{j+1}(\rho_{j}(\dt))$, using \eqref{LPVILC:eq:ParameterUpdate}.
			\end{enumerate}
		\end{enumerate}
		\hrule
	\end{proced}
\end{figure}
\section{Experimental Setup Characterization}
In this section, the experimental setup is introduced and characterized, leading to an appropriate parameter-varying feedforward parameterization, hence contributing to C3. 
Specifically, several preliminary experiments are performed to determine which feedforward structure and kernels will be used. 
\subsection{Experimental Setup}
	\label{LPVILC:sec:expSetup}
\begin{figure}[tb]
		\centering
		\includegraphics{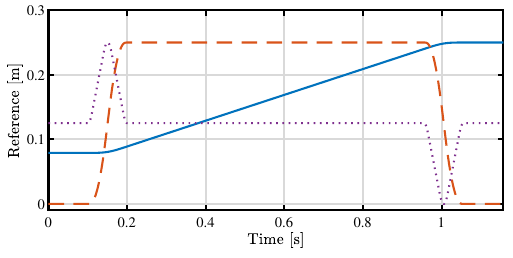}
		\caption{Reference $r$ \markerline{mblue}, scaled reference velocity $\dot{r}$ \markerline{mred}[densely dashed], and scaled reference acceleration \markerline{mpurple}[dotted] applied to the experimental setup.}
		\label{LPVILC:fig:ExperimentalReference}
	\end{figure}

	The experimental setup considered is a 1 degree-of-freedom belt-driven carriage, which is mounted on a linear guide, as shown in \secRef{LPVILC:sec:MotivatingApplication} and \figRef{LPVILC:fig:ExperimentalSystem}. The BLAC motor is equipped with a rotary encoder having a resolution of 8192 counts per revolution, in addition to a linear encoder on the linear guide with a resolution of 100 nm. The timing belt has a tooth pitch of 2 mm, which is made of rubber with an added carbon fiber core for additional stiffness. The pulleys have 15 teeth per revolution, and hence, a circumference of 0.03 m. The distance between the pulleys is $L=0.3$ m. The goal is to track a third-order scanning motion with the carriage using the input to the motor $u$, consisting of a large constant velocity part, which is seen in \figRef{LPVILC:fig:ExperimentalReference}.
	The performance is evaluated during constant velocity, since this type of drivetrain is typically used in scanning applications such as printing systems.
	
	The control scheme in \figRef{LPVILC:fig:FBLayout} is used, where the LTI feedback controller is a discrete-time lead filter with an additional low-pass filter described by the transfer function
	\begin{equation}
		\label{LPVILC:eq:ExperimentalController}
		\begin{aligned}
			\mathcal{C}(q)= \frac{2169 q^{-1} - 381 q^{-2} - 1747 q^{-3}}{1 - 2.421 q^{-1} + 1.961 q^{-2} - 0.5337 q^{-3}}.
		\end{aligned}
	\end{equation}
		The settings during experimentation are shown in \tabRef{LPVILC:tab:expValues}.
		\begin{table}[H]
			\centering
			\caption{Experimental settings.}
			\label{LPVILC:tab:expValues}
			\begin{tabular}{llll}
				\toprule
				\textbf{Variable}    & \textbf{Abbrevation} & \textbf{Value} & \textbf{Unit} \\
				\midrule
				Sampling time   & $T_s$            & $2.5\cdot10^{-4}$       & s   \\
				Number of samples    & N               & 4630    & -    \\
				Reference stroke & - & 0.171 & m \\
				Maximum velocity & - & 0.2 & m/s \\
				Maximum acceleration & - & 4 & m/s$^2$ \\
				\bottomrule
			\end{tabular}
		\end{table}
\subsection{Characterization of Position-Dependent Dynamics}
	\label{LPVILC:sec:charPosDep}
\begin{figure}[tb]
\centering
\includegraphics{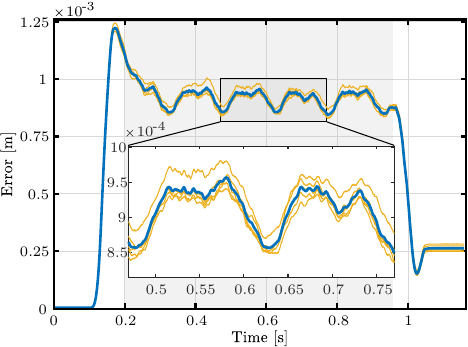}
\caption{During constant velocity \markerline{gray}[solid][x][0][6][8pt][0.2][0.2], the experimental tracking error for 5 repetitions of the reference in \figRef{LPVILC:fig:ExperimentalReference} with zero feedforward $f_j=0$ \markerline{myel}[solid][x][0][0.5] and their sample mean \markerline{mblue} show highly repeatable position-dependent effects.}
\label{LPVILC:fig:1DOFSlider_PrelimTracking}
\end{figure}

	The tracking error with zero feedforward $f_j=0$ for 5 repetitions of the reference in \figRef{LPVILC:fig:ExperimentalReference} is shown in \figRef{LPVILC:fig:1DOFSlider_PrelimTracking}.
The following is observed from the tracking error in \figRef{LPVILC:fig:1DOFSlider_PrelimTracking}.
\begin{itemize}
\item The tracking error is highly repeatable, and hence, an iterative approach is suitable.
\item The large offset in the tracking error indicates that both Coulomb and viscous friction feedforward might be necessary.
\item During acceleration, the error reaches its maximum, motivating the need for acceleration feedforward.
\item The tracking error has a period of 0.15 s during a constant velocity of 0.2 m/s, resulting in a spatial period of 0.03 m, which is the circumference of the pulley.
\end{itemize}

\begin{figure}[tb]
\centering
\includegraphics{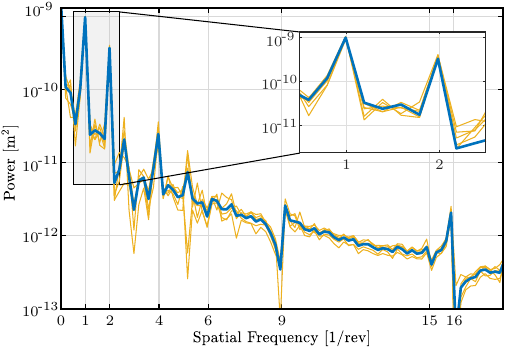}
\caption{Power spectrum of the tracking error during constant velocity for 5 times tracking the reference in \figRef{LPVILC:fig:ExperimentalReference} with zero feedforward $f_j=0$ \markerline{myel}[solid][^][0][0.5] and their sample mean \markerline{mblue} shows highly repetitive behavior in spatial domain.}
\label{LPVILC:fig:PSD_NOFF}
\end{figure}

The spatial periodic effect is further analyzed with a power spectrum of the tracking error during constant velocity as a function of the number of pulley revolutions in \figRef{LPVILC:fig:PSD_NOFF}.
The spatial power spectrum shows that during constant velocity, the error is dominated by the zero frequency, the fundamental frequency [1/rev], its second harmonic [2/rev] and marginally by its fourth harmonic [4/rev]. The zero frequency contribution is mainly caused by the lack of Coulomb friction feedforward, which is also seen from the constant offset of the tracking error in \figRef{LPVILC:fig:1DOFSlider_PrelimTracking}. The fundamental frequency, and second and fourth harmonic are most likely caused due position-dependent effects introduced by pulley out-of-roundness or motor cogging.
\section{Experimental Application}
In this section, both LTI feedforward control and the developed LPV feedforward control methods are compared on the experimental setup, further contributing to C3. 
Both the learning procedure and the tracking results are shown.
\subsection{Experimental Learning Settings}
\begin{figure}[tb]
		\centering
		\includegraphics{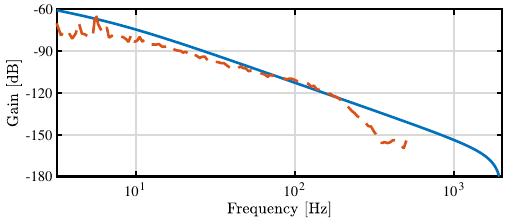}
		\caption{The LTI model $\widehat{\mathcal{G}}(q)$ \markerline{mblue} is constructed to approximate the measured frequency response function \markerline{mred}[densely dashed] of the experimental belt-driven carriage.}
		\label{LPVILC:fig:ExpFRF}
	\end{figure}

	The model of the process sensitivity $\widehat{\mathcal{J}}$ is determined by using a simplified LTI model of the system $\mathcal{G}$ in feedback with the controller $\mathcal{C}$ in \eqref{LPVILC:eq:ExperimentalController}. The simplified LTI model $\widehat{\mathcal{G}}$ is determined by using the measured frequency response function in \figRef{LPVILC:fig:ExpFRF}.
	As a result, the simplified LTI model $\widehat{\mathcal{G}}$ consists of a mass with a damper attached to the fixed world, and is discretized with a zero order hold, resulting in
	\begin{equation}
		\begin{aligned}
			\widehat{\mathcal{G}}(q) =\frac{2.82 q^{-1} + 2.798 q^{-2}}{1 - 1.978 q^{-1} + 0.9775 q^{-2}}\cdot10^{-8}.
		\end{aligned}
	\end{equation}
	After interconnecting $\widehat{\mathcal{G}}$ with the controller $\mathcal{C}$ from \eqref{LPVILC:eq:ExperimentalController}, the LTI model of the process sensitivity is
	\begin{equation}
		\label{LPVILC:eq:ExperimentalProcessSensitivity}
		\resizebox{\linewidth}{!}{$\displaystyle
		\begin{aligned}
			\widehat{\mathcal{J}}(q) = \frac{ 2.82 q^{\smallMin1}\!- 4.029 q^{\smallMin2} \!- 1.244 q^{\smallMin3} \!+ 3.982 q^{\smallMin4} \!- 1.493 q^{\smallMin5}}{1\! -\! 4.399 q^{\smallMin 1}\!\! +\! 7.727 q^{\smallMin2}\! - 6.779 q^{\smallMin3}\! + 2.972 q^{\smallMin4}\! - \!0.522 q^{\smallMin 5}}\!\cdot\!10^{\smallMin8}\!.
		\end{aligned}
		$}
	\end{equation}
	
	The reference tracking task is constant during learning $r_j=r \;\forall j$ and the same as in \secRef{LPVILC:sec:expSetup}, shown in \figRef{LPVILC:fig:ExperimentalReference}. The reference signal is used as scheduling sequence $\rho_j=r \; \forall j$, since this is relatively close to the position of the carriage $y$, and it is known in advance of the tracking experiment. 
	
	The observations in \secRef{LPVILC:sec:charPosDep} motivate using viscous friction, Coulomb friction, and acceleration feedforward in \eqref{LPVILC:eq:FFLaw}, which is widely applied in feedforward control \citep{Steinbuch2010}, i.e.,
	\begin{equation}
		\label{LPVILC:eq:ExperimentalLPVFFStructure}
		\begin{aligned}
			f_j\left( \dt\right)  = \theta^a_j\left( \rho\left( \dt\right) \right)  \ddot{r}\left( \dt\right)+\theta^v_j\left( \rho\left( \dt\right) \right)  \dot{r}\left( \dt\right)+\theta^c_j\left( \rho\left( \dt\right) \right)  \operatorname{sign}\left(\dot{r}\left( \dt\right)\right),
		\end{aligned}
	\end{equation}
	with discrete-time derivatives $\ddot{r}(\dt)$ and $\dot{r}(\dt)$. The corresponding basis functions and feedforward parameters from \eqref{LPVILC:eq:FFLaw} are
	\begin{equation}
		\begin{aligned}
			\psi\left( r\left( \dt\right) \right)  &= \begin{bmatrix}
				\left(\frac{q-q^{-1}}{2T_s}\right)^2 r(\dt)& \frac{q-q^{-1}}{2T_s} r(\dt) & \operatorname{sign}\left(\frac{q-q^{-1}}{2T_s} r(\dt)\right)
			\end{bmatrix}, \\
			\theta_j\left( \rho\left( \dt\right) \right) &= \begin{bmatrix}
				\theta^a_j\left( \rho\left( \dt\right) \right) & \theta^v_j\left( \rho\left( \dt\right) \right)  & \theta^c_j\left( \rho\left( \dt\right) \right)
			\end{bmatrix}^\top,
		\end{aligned}
	\end{equation}
	where discrete-time derivatives $\ddot{r}(\dt)$ and $\dot{r}(\dt)$ from \eqref{LPVILC:eq:ExperimentalLPVFFStructure} are computed using the central difference. The $3\times3$ kernel function matrix \eqref{LPVILC:eq:kernelfunctionmatrix} is defined such that the estimated feedforward parameters $\theta_j\left( \rho\left( \dt\right) \right)$ do not correlate, and the acceleration and Coulomb friction feedforward parameter functions are enforced to be constant using the kernel \eqref{LPVILC:eq:constantKernel}, i.e.,
	\begin{equation}
		\label{LPVILC:eq:ExperimentalKernelFunctionMatrix}
		K\left(\rho^*,\rho\right) = \begin{bmatrix}
			k^c\left(\rho^*,\rho\right) & 0 & 0 \\
			0 & k_{22}\left(\rho^*,\rho\right) & 0  \\
			0 & 0 & k^c\left(\rho^*,\rho\right) 
		\end{bmatrix}.
	\end{equation}
	The hyperparameters of the constant kernels for the acceleration and coulomb friction feedforward parameter functions are respectively $\sigma^2=1$ and $\sigma^2=3$. The amount of trials is set to 10 and the regularization coefficient $\gamma$ from \eqref{LPVILC:eq:RegularizedCostFunction} to $\gamma=5\cdot10^{-5}$. 
	
\begin{figure}[tb]
\centering
\includegraphics[width=0.8\linewidth]{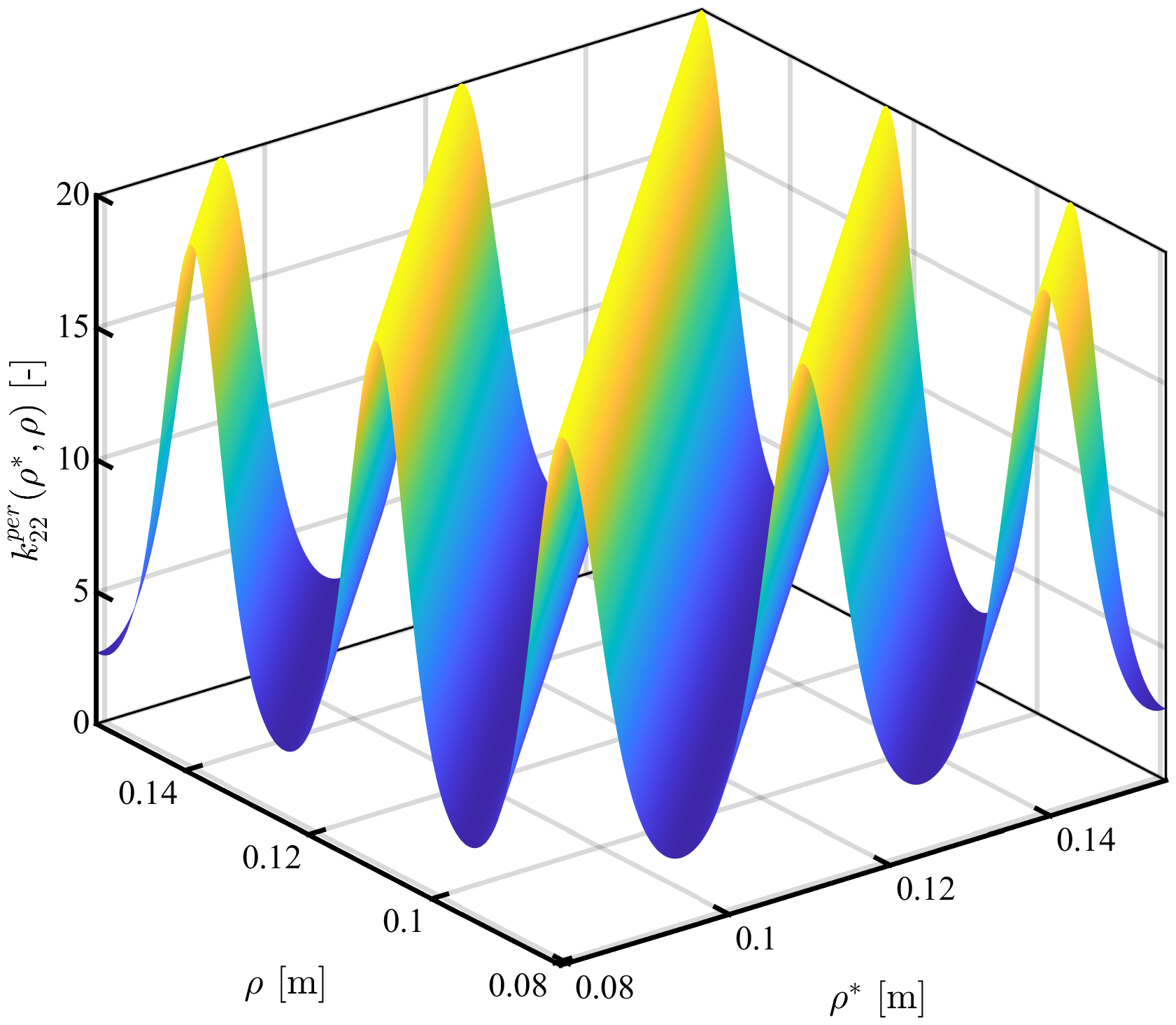}
\caption{Surface plot of kernel function $k_{22}^{per}(\rho^*,\rho)$ evaluated at a scheduling sequence $\rho^*\!,\rho \in \mathbb{R}_{[0.08,0.16]}$ \surfaceLegendNormal{}, which enforces a smooth and periodic function having a period of $p=0.03$ m.}
\label{LPVILC:fig:1DOFSlider_KernelSurf}
\end{figure}

LTI feedforward control and the developed LPV feedforward control are compared by using a constant or varying viscous friction feedforward parameter function. Specifically, the compared methods use the following kernel functions for the viscous friction feedforward parameter.
\begin{itemize}
\item \textbf{LTI feedforward}: constant kernel from \eqref{LPVILC:eq:constantKernel}, $k_{22}\left( \rho^*,\rho\right)=k^c\left( \rho^*,\rho\right)$, with hyperparameter $\sigma^2 = 20$.
\item \textbf{LPV feedforward}: periodic kernel from \eqref{LPVILC:eq:per}, $k_{22}\left( \rho^*,\rho\right)=k^{per}_{22}\left( \rho^*,\rho\right),$ with hyperparameters $\sigma^2 = 20$, $\ell = 1$ m and $p = 3\cdot 10^{-2}$ m.
\end{itemize}
Note that the LTI feedforward approach recovers ILC with basis functions, where the feedforward parameters are estimated using Tikhonov regularization, see \remRef{rem:RegularizedLTIBFILC}. The spatial periodic effect of the tracking error in \secRef{LPVILC:sec:charPosDep} motivates to select $k_{22}$ being periodic in the pulley circumference. A surface plot of the kernel function $k^{per}_{22}$ is shown in \figRef{LPVILC:fig:1DOFSlider_KernelSurf}. LTI feedforward control recovers LTI ILC with basis functions \citep{VanDerMeulen2008}, where Tikhonov regularization is used to estimate the feedforward parameters.
\subsection{Experimental Results}
\begin{figure}[tb]
\centering
\includegraphics[width = 0.99 \linewidth]{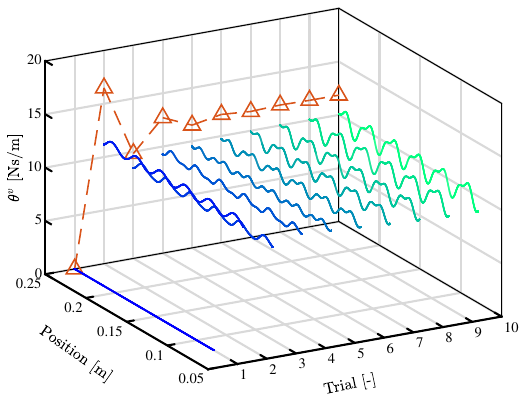}
\caption{Feedforward parameter function $\theta_j^v(\rho(\dt))$ over the trials estimated by the developed LPV feedforward method \surfaceLegend{} and LTI feedforward method $\theta_j^v$ \markerline{mred}[densely dashed][triangle][3][0.75].}
\label{LPVILC:fig:ExperimentalLPVFFParameter}
\end{figure}

\begin{figure}[tb]
\centering
\includegraphics{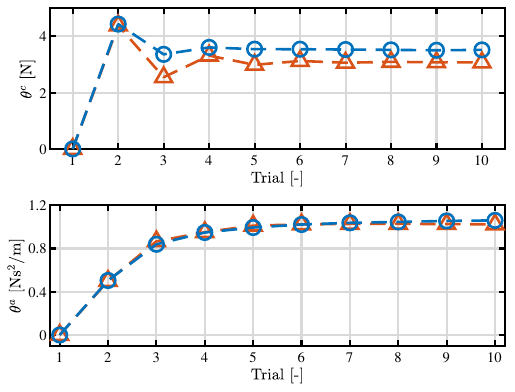}
\caption{Constant feedforward parameters $\theta^c_j$ and $\theta^a_j$ estimated by LTI feedforward control \markerline{mred}[densely dashed][triangle][3][0.75] and LPV feedforward control \markerline{mblue}[densely dashed][o][2.5][0.75].}
\label{LPVILC:fig:ExperimentalLTIFFParameter}
\end{figure}

The learned viscous, Coulomb, and acceleration feedforward parameter functions are shown in \figRef{LPVILC:fig:ExperimentalLPVFFParameter} and \figRef{LPVILC:fig:ExperimentalLTIFFParameter}, which demonstrate that the learning has converged.
The viscous friction feedforward parameter function clearly shows the periodic behavior enforced by the periodic kernel. The Coulomb friction and acceleration feedforward parameters do not show any significant difference between the LTI and LPV feedforward methods.

\begin{figure}[tb]
\centering
\includegraphics{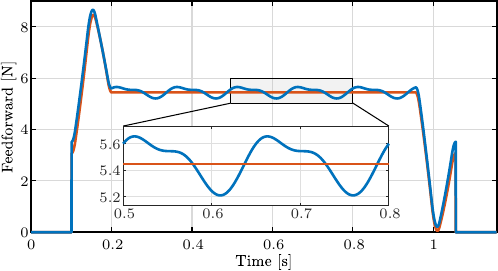}
\caption{The feedforward signal during the final trial $f_{10}$ shows the periodic effect obtained by the LPV feedforward signal \markerline{mblue} compared to the LTI feedforward signal \markerline{mred}.}
\label{LPVILC:fig:ExperimentalTFeedforward}
\end{figure}

The converged feedforward signal in the final trial from \eqref{LPVILC:eq:ExperimentalLPVFFStructure} using the feedforward parameters from \figRef{LPVILC:fig:ExperimentalLPVFFParameter} and \figRef{LPVILC:fig:ExperimentalLTIFFParameter} for the reference in \figRef{LPVILC:fig:ExperimentalReference} is shown in \figRef{LPVILC:fig:ExperimentalTFeedforward}.

\begin{figure}[tb]
\centering
\includegraphics{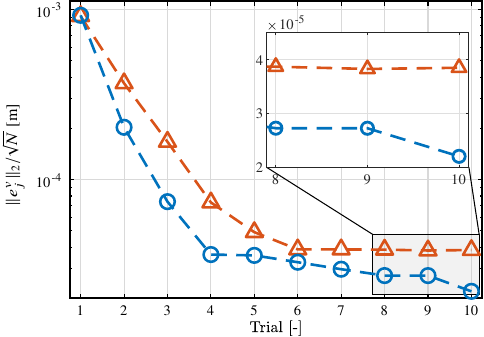}
\caption{The RMS error during constant velocity $e_j^v$ converges faster and to a lower value for the developed iterative kernel-regularized estimates of LPV parameters \markerline{mblue}[densely dashed][o][2.5][0.75] compared to the LTI feedforward parameters \markerline{mred}[densely dashed][triangle][3][0.75].}
\label{LPVILC:fig:1DOFSlider_RMSError}
\end{figure}

\begin{figure}[tb]
\centering
\includegraphics{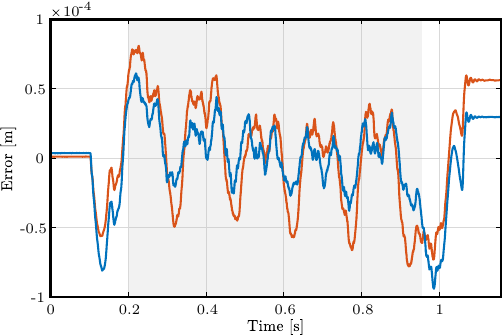}
\caption{During constant velocity \markerline{gray}[solid][x][0][6][8pt][0.2][0.2], the tracking error in the final trial $e_{10}$ is reduced by the developed iterative kernel-regularized estimates of LPV feedforward parameters \markerline{mblue} compared to the LTI feedforward parameters \markerline{mred}.} 
\label{LPVILC:fig:ExperimentalTrackingError}
\end{figure}

\begin{figure}[tb]
\centering
\includegraphics{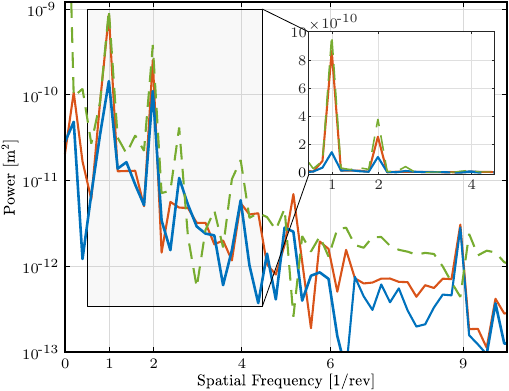}
\caption{The power of the tracking error for the final trial $e_{10}$ is significantly lower for the developed LPV feedforward control \markerline{mblue} compared to LTI feedforward control \markerline{mred} and zero feedforward $f_j=0$ \markerline{mgreen}[densely dashed].}
\label{LPVILC:fig:ExperimentalErrorSpectra}
\end{figure}

The converged reference tracking performance is illustrated through the Root-Mean-Squared (RMS) tracking errors in \figRef{LPVILC:fig:1DOFSlider_RMSError}, the tracking error in the final trial in \figRef{LPVILC:fig:ExperimentalTrackingError}, and the power spectrum of the final tracking error in \figRef{LPVILC:fig:ExperimentalErrorSpectra}. In addition, several error metrics that are particularly relevant for systems performing scanning motions, including the RMS and maximum value of the error during constant velocity $e_j^v$, are presented in \tabRef{LPVILC:tab:ExperimentalErrorMetric}. The following observations are made concerning the tracking performance.
\begin{table}[tb]
\centering
\caption{Experimental error metrics during constant velocity for LTI and LPV feedforward.}\label{LPVILC:tab:ExperimentalErrorMetric}
\begin{tabular}{llll}
\toprule
\textbf{Metric} & \textbf{LTI} & \textbf{LPV} & \textbf{Unit} \\ \midrule
$\|e_{10}^v\|_2/\sqrt{N}$  & $3.85\cdot10^{-5}$ &  \textbf{ $2.2\cdot10^{-5}$ } & [m]\\
$\|e_{10}^v\|_\infty$  & $8.1\cdot10^{-5}$  & $6.1\cdot10^{-5}$  & [m]\\
Power $e_{10}^v$ (1/rev) & $8.65\cdot10^{-10}$    & $1.44\cdot10^{-10}$ & [m$^2$]  \\
Power $e_{10}^v$ (2/rev) & $2.54\cdot10^{-10}$ & $1.10\cdot10^{-10}$ &[m$^2$]  \\ \bottomrule                   
\end{tabular}
\end{table}
\begin{itemize}
\item The RMS error during constant velocity in the final trial is 43 \% lower for LPV feedforward than for LTI feedforward as shown in \figRef{LPVILC:fig:1DOFSlider_RMSError} and \tabRef{LPVILC:tab:ExperimentalErrorMetric}.
\item The maximum absolute error in the final trial $\left\|e_{10}\right\|_\infty$ during constant velocity improved by 25 \% by using LPV feedforward compared to LTI feedforward, as illustrated in \figRef{LPVILC:fig:ExperimentalTrackingError} and \tabRef{LPVILC:tab:ExperimentalErrorMetric}.
\item The amplitude of the first and second harmonic are decreased by respectively a factor 6 and 2.3 in terms of their power, as shown in \figRef{LPVILC:fig:ExperimentalErrorSpectra} and \tabRef{LPVILC:tab:ExperimentalErrorMetric}.
\end{itemize}
The observations show that learning feedforward parameter functions using the developed iterative kernel-regularized estimator increases performance significantly for motion systems.
\section{Conclusions}
The results in this \manuscript enable data-driven learning of parameter-variations in feedforward control, which can significantly improve the tracking performance of LPV systems. The key idea is to iteratively learn LPV feedforward parameter functions using kernel-regularized estimation. Kernel-regularized function estimation is advantageous since it is non-parametric, meaning no specific structure needs to be enforced on the dependency on the scheduling sequence. The iterative learning approach directly optimizes the tracking error, which enhances the estimation quality without the need for accurate system models. The developed framework is experimentally validated on a belt-driven motion system, demonstrating effective compensation of position-dependent drivetrain dynamics. The developed method demonstrates significant potential for industrial applications, particularly in mechatronic systems, by enabling improved tracking performance.

Future work includes experimental comparison of the developed framework to other LPV or non-linear feedforward control approaches.
\bibliographystyle{elsarticle-num}
\bibliography{../../library}
\vspace{2mm}
\setlength\intextsep{0pt}
\footnotesize
\begin{wrapfigure}{l}{24mm} 
	\includegraphics[width=24mm,clip,keepaspectratio]{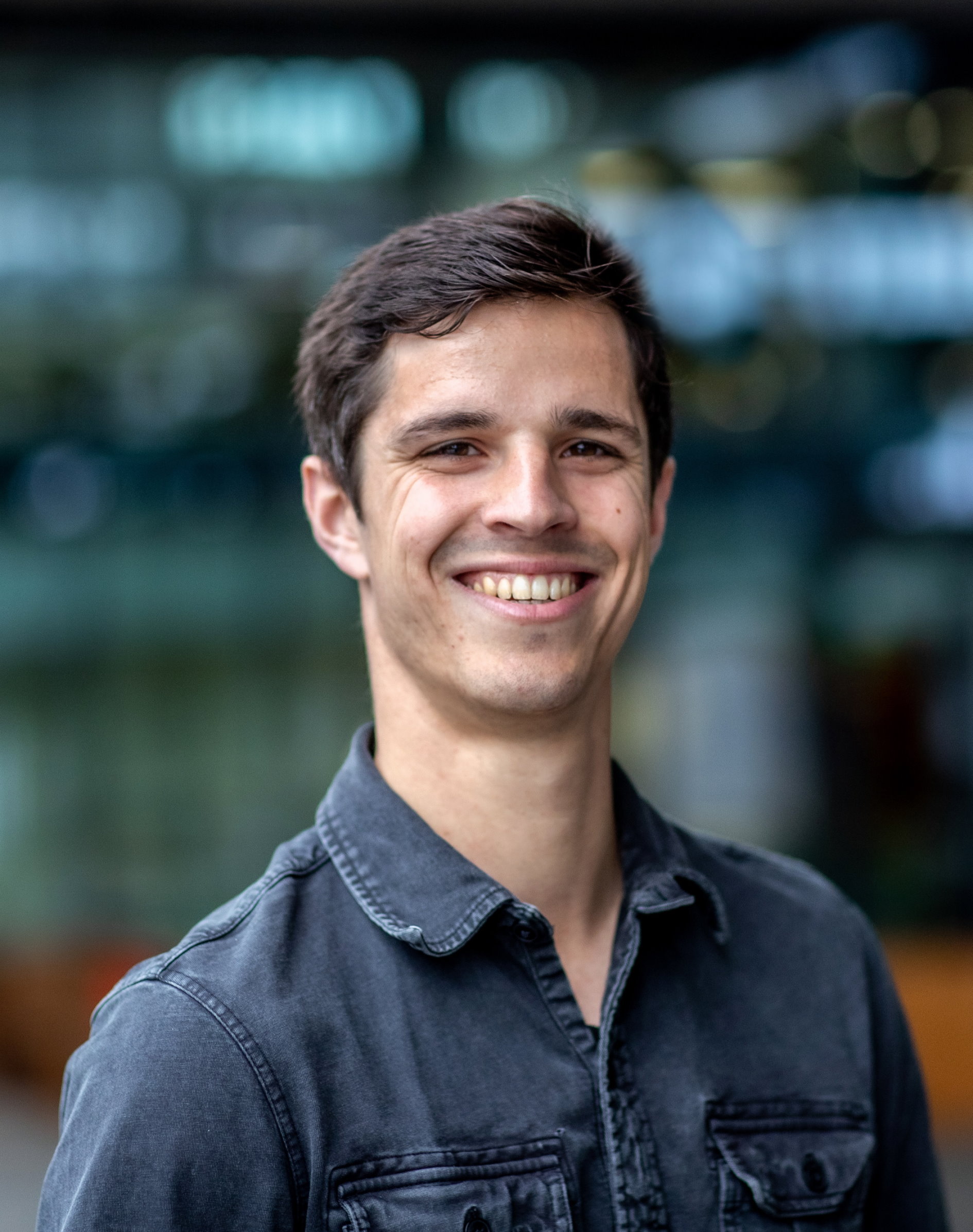}
\end{wrapfigure} \par \noindent
\textbf{Max van Haren} is currently working toward the Ph.D. degree in mechanical engineering with the Control Systems Technology section, Eindhoven University of Technology. He received the M.Sc. degree (cum laude) in mechanical engineering in 2021 from the Eindhoven University of Technology, Eindhoven, The Netherlands. His research interests include control and identification of mechatronic systems, including sampled-data, multirate, and linear parameter-varying systems. \par \vspace{1mm}
\begin{wrapfigure}{l}{24mm} 
	\includegraphics[width=24mm,clip,keepaspectratio]{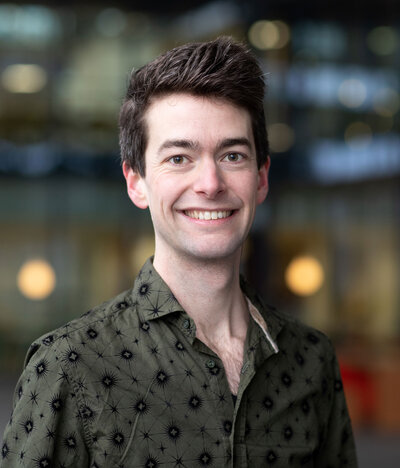}
\end{wrapfigure}\par \noindent
\textbf{Lennart Blanken} is currently a System Designer Mechatronics with Sioux Technologies, Eindhoven, The Netherlands. Additionally, he is an assistant professor with the Department of Mechanical Engineering at the Eindhoven University of Technology. He received the M.Sc. degree (cum laude) and Ph.D. degree in mechanical engineering from the Eindhoven University of Technology, Eindhoven, The Netherlands, in 2015 and 2019, respectively. His research interests include advanced feedforward control, learning control, repetitive control, and their applications to mechatronic systems. \par \vspace{1mm}
\begin{wrapfigure}{l}{24mm} 
	\includegraphics[width=24mm,clip,keepaspectratio]{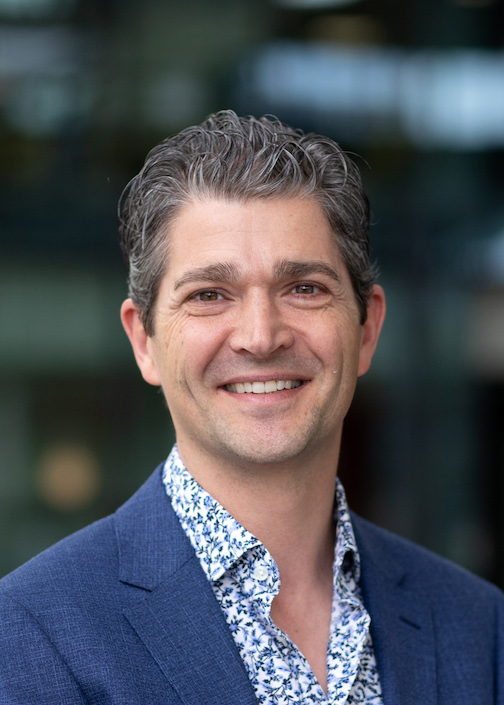}
\end{wrapfigure}\par \noindent
\textbf{Tom Oomen} is full professor with the Department of Mechanical Engineering at the Eindhoven University of Technology. He is also a part-time full professor with the Delft University of Technology. He received the M.Sc. degree (cum laude) and Ph.D. degree from the Eindhoven University of Technology, Eindhoven, The Netherlands. He held visiting positions at KTH, Stockholm, Sweden, and at The University of Newcastle, Australia. He is a recipient of the 7th Grand Nagamori Award, the Corus Young Talent Graduation Award, the IFAC 2019 TC 4.2 Mechatronics Young Research Award, the 2015 IEEE Transactions on Control Systems Technology Outstanding Paper Award, the 2017 IFAC Mechatronics Best Paper Award, the 2019 IEEJ Journal of Industry Applications Best Paper Award, and recipient of a Veni and Vidi personal grant. He is currently a Senior Editor of IEEE Control Systems Letters (L-CSS) and Co-Editor-in-Chief of IFAC Mechatronics, and he has served on the editorial board of IEEE Transactions on Control Systems Technology. He has also been vice-chair for IFAC TC 4.2 and a member of the Eindhoven Young Academy of Engineering. His research interests are in the field of data-driven modeling, learning, and control, with applications in precision mechatronics. \par


\end{document}